\documentclass{elsart}
\usepackage{graphicx}
\usepackage{amssymb}
\begin{document}
\begin{frontmatter}
\title{The use of cosmic muons in detecting heterogeneities in large
volumes}
\author[mx]{V. Grabski\corauthref{cor}}
\ead{varlen.grabski@cern.ch,grabski@fisica.unam.mx},
\corauth[cor]{Corresponding author. Tel: 52-555-622-5185; Fax: 52-555-622-5009}
\author[fr]{R. Reche},
\author[mx]{R. Alfaro},
\author[mx]{E. Belmont},
\author[mx]{A. Mart\'inez-D\'avalos},
\author[mx]{A. Sandoval} and
\author[mx]{A. Menchaca-Rocha}
\ead{menchaca@fisica.unam.mx}
\address[mx]{Departamento de F\'isica Experimental, Instituto de F\'isica
\newline \centerline{Universidad Nacinal Aut\'onoma de M\'exico.}
\newline \centerline{M\'exico, D.F. C.P. 04510}}
\address[fr]{Laboratoire de Physique Subatomique et de Cosmologie, Grenoble, France}

\begin{abstract}
The muon intensity attenuation method to detect heterogeneities in large matter volumes is analyzed. Approximate analytical expressions to estimate the collection time and the signal to noise ratio, are proposed and validated by Monte Carlo simulations. Important parameters, including point spread function and coordinate reconstruction uncertainty are also estimated using Monte Carlo simulations.

\end{abstract}

\end{frontmatter}

\section{Introduction}

The first practical application of cosmic muon intensity attenuation measurements dates back 60 years, when attempts were made to evaluate snow layers in Australian mountains \cite{Georg}. Ten years later Alvarez et al. \cite{Alvarez} used this principle to search for hidden chambers in the great Egyptian pyramid of Cheops, in Giza. The use of near-horizontal muon attenuation to probe active volcanoes \cite{Nagamine,Tanaka1,Tanaka2} or to detect and locate heavy metals in border-crossing vehicles using scattering muon tomography\cite{Borozdin,Schultz}, which can also be combined with muon directional intensity attenuation, electro-magnetic shower induction, and muonic X-ray emission as additional means for the non-destructive assay of small and medium size objects \cite{Klimenko}, or the use of muon-induced neutron emission to gauge carbon liner thickness in steel industry blast furnaces\cite{Gilboy}, are good examples of the continued interest in applying cosmic muon detection to a wide variety of practical problems. 

In our previous works \cite{Alfaro,Grabski1}, we explored the viability of carrying out an experiment similar to that of Alvarez et al.\cite{Alvarez} in the Pyramid of the Sun at Teotihuacan, in Central Mexico. Here we evaluate important experimental and simulation planning parameters, such as detector size, coordinate resolution and point spread function (PSF), the signal-to-noise ratio and observation time to detect a structure of a given size. Some of those parameters are estimated analytically using an approximate description of muon directional intensity (DI). Here we shall concentrate on an energy range relevant to explore matter depths in the range $5-50$ $ kg/cm^{2}$.
Monte Carlo simulations have also been performed to estimate the magnitude of PSF and the coordinate resolution as functions of the matter thickness along the muon track, as well as to test the approximate expression proposed below for the signal to noise ratio.

\section{Muon energy spectrum and signal to noise ratio}

The muon DI attenuation method \cite{Georg} requires a knowledge of the initial cosmic muon differential energy spectrum at the site of interest $F(E,\Theta)$, where $E$ is the muon energy and  $\Theta$ the zenith angle. For the vertical direction ($\Theta=0$) at sea level $F$ can be represented by \cite{Grieder}:

\begin{eqnarray}
F(E,\Theta=0) & = & G E^{-n(E)} [cm^{-2}s^{-1}sr^{-1}GeV^{-1}],\label{eq:1}
\end{eqnarray}

where $G$, is a normalization constant and the spectral index $n(E)$ is a slowly varying energy-dependent function. A well established\cite{Gaiser,CERN} reference value for the spectral index is $n (E = 100$ $GeV) = 2.7$. This fact can be used to extract the value of $G$ at that point. The energy dependence of $n(E)$ can then be determined using the experimental data. 

Muon differential energy spectrum measurements exist with sufficient accuracy at sea level and within the energy range $E < 100$ $GeV$ \cite{Hebecker}. These data have been compiled and analyzed by various authors\cite{Gaiser,CERN,Hebecker,Bedevi}, resulting in a variety of parameterizations. Among them the one by  Hebbeker and Timerman \cite{Hebecker} describes the data in the energy  region of $10 - 100$ $GeV$ with an accuracy better than $2\%$, while for $E > 100$ $GeV$ the experimental data has larger uncertainties, of the order of $10\%-15\%$. The results of $n(E)$ obtained using the parameterizations \cite{Hebecker} and \cite{Gaiser,CERN} is represented in Fig. 1.

The DI attenuation method \cite{Georg} is based on using the muon integral energy spectrum $\Phi(E>E_{L},\Theta)$ (where $E_{L}$ is the muon mean threshold energy  to cross the volume of interest expressed in GeV, for a matter thickness $L$ along the muon track, given in density length units\cite{Nagamine}), defined as \cite{Grieder}:

\begin{eqnarray}
\Phi(E_{L},\Theta=0) & = & \int_{E_{L}}^{\infty}F(E,\Theta=0)dE.\label{eq:2}
\end{eqnarray}

As shown in Fig 1 (solid line), $n(E)$ varies less than $\pm 1.5\%$ in the energy region $E = 10 - 100$ $GeV$, so that an $n(E) = n_{c} = constant$ approximation can be  used to obtain a simplified analytic expressions for experiment planning purposes.  Using $n_{c} = 2.687$ provides (see Fig 2) a $\sim10$\% accuracy for both differential and integral vertical energy spectra.

Locating an heterogeneity in the matter volume using muon intensity attenuation is based on calculating the difference $\Delta N$ between the experimental number of muons $N$ (after crossing the volume of interest) and the simulated number of muons in an equivalent volume in which no heterogeneity is included. Hereafter we shall refer to this as the "simulated background". If $\sigma_{\Delta N}$ represents the uncertainty of $\Delta N$, the signal to noise ratio, ($\xi$) is defined as:

\begin{equation}
\xi\equiv\frac{\Delta N}{\sigma_{\Delta N}}.\label{eq:3}
\end{equation}

Note that the uncertainty associated with the simulated background can allways be reduced below the experimental one, thus $\sigma_{\Delta N} \approx \sqrt{N}$.

Let us address the question of what is best to measure, $F$ or $\Phi$. Within $n(E) = n_{c}$ approximation,  one can calculate the time-integrated vertical muon count $N$ and the difference $\Delta N$ for an energy $E_L$, and the energy difference $\Delta E$ associated to the heterogeneity, registered in a detector having surface ${S}$, solid angle acceptance $\Omega$ and during collection time ${T}$ using $N=G\Phi(E_{L},\Theta=0)\Omega S T$ and $\Delta N \approx GF(E_{L},\Theta=0)\Delta E\Omega S T$ (assuming that $\Delta E \ll E_{L}$). Thus, Eq.(\ref{eq:3}) can be used to estimate the signal to noise ratio ($\xi_\Phi$)  associated with the integral intensity $\Phi$. A similar procedure can be used to calculate the signal to noise ratio ($\xi_{F}$) associated with a finite energy interval $\varepsilon \ll E_{L}$ using $F$. Then the ratio between $\xi_{\Phi}$ and $\xi_{F}$ can be written approximately as:

\begin{equation}
\xi_{\Phi}/\xi_{F}\approx\sqrt{\frac{E_{L}(n_{c}-1)}{\varepsilon n_{c}^{2}}},\label{eq:4}
\end{equation}

From this expression it is clear that it becomes preferable to measure the integral intensity (rather than the differential one) when $E_{L} > \varepsilon n_{c}^{2}/( n_{c} -1) \approx 4.28  \varepsilon$. Since the $n(E)=n_{c}$ approximation is valid for $E > 10 GeV$, considering $ \varepsilon = 1$ $GeV$ (to fulfill $\varepsilon \ll E_{L}$) implies $E_{L} > 4.28$ $GeV$ for the $\xi_{\Phi}>\xi_{F}$ condition to be fulfilled. Still, the use of differential energy spectrum gives a better ratio for $\Delta N/N$ \cite{Alvarez}. This consequence of the power-law shape of the muon energy spectra justifies the important experimental simplification of measuring $\Phi$ rather than $F$. Thus, from now on we consider only the signal to noise ratio for the integral energy intensity.

Taking into account that there is a unique relationship between a moun energy and the mean range, $\xi$ can also be expressed as a function of $L$, which is more convenient in some practical applications. For standard soil matter, with average density $2\times 10^{-3}$ $kg/cm^{3}$, and within the energy interval $5-100$ $GeV$ the calculated data for the range\cite{Theriot} can be fitted using ${E_{L}=cL^{k}}$\footnote{This is an overestimation of $E_{L}$, because the steepness of $F\propto E^{-2.7}$ leads to an asymetric distortion, an enhancement of the low energy contribution. Hear we neglect this effect due to the narow energy distribution of muons  having a given range value $L$}, where ${L}$ is expressed in ${kg/cm^{2}}$, $c=1.84$ $GeV/(kg/cm^{2})^{k}$ and ${k = 1.074}$.

When the localized density heterogeneity has a size $\Delta L\ll L$, and still within the $n(E) = n_c$ approximation $\xi$ can be expressed as a function of $\Delta L$ and $L$ as follows:

\begin{equation}
\xi  \approx  \sqrt{GST\Omega(n_{c}-1)}\frac{ k\Delta L}{c^{(n_{c}-1)/2}L^{k(n_{c}-1)/2+1}},\label{eq:5}
\end{equation}

This implies that, for a fixed heterogeneity size $\Delta L$, $\xi$ decreases as $L^{-1.906}$. To illustrate the impact of our approximation on $\xi$, the ratio $\xi_{app}/\xi_{ex}$ between $\xi$ corresponding to the approximate ($\xi_{app}$) and the exact ($\xi_{ex}$) differential muon intensities, for different $\Delta L/L$ values, as a function of $L$ is presented in Fig 3. There is an agreement within $10$\% for $\Delta L/L<10$\% in the range $L = 8-50$ $kg/cm^{2}$.

Concerning the experimental statistics necessary to achieve a given $\xi$, still within the constant spectral index $n(E)$ approximation, we find:

\begin{equation}
N\thickapprox(\xi\frac{L}{k(n_{c}-1)\Delta L})^{2}.\label{eq:6}
\end{equation}

Thus, the required collection time for a given $\xi$ can be estimated using:

\begin{eqnarray}
T & \approx & \frac{\xi^{2} c^{n_{c}-1}L^{k(n_{c}-1)+2}}{G\Omega S (n_{c}-1)k^{2}\Delta L^{2}} .\label{eq:7}
\end{eqnarray}

Since the collection time increases as $L^{3.812}$, to minimize $T$ it is necessary to choose a detector location where $L$ gets its minimum possible value for the region of interest.

Using the value of $\Delta L$, on can estimate the geometrical length $\Delta X$ of the heterogeneity, measured along the muon direction as:

\begin{equation}
\Delta X=\Delta L/\rho,\label{eq:8}
\end{equation}

where $\rho$ is the average matter density, expressed in $kg/cm^{3}$. Assuming that the density heterogeneity length $\Delta X\ll X$, where $X$ is the total length of the muon trajectory inside 
the volume of interest, the uncertainty of $\Delta X$, $\sigma_{\Delta X}$ can be estimated from:

\begin{equation}
\frac{\sigma_{\Delta X}}{\Delta X}\approx\sqrt{(\frac{1}\xi)^{2}+(\frac{\sigma_{F}}{F})^{2}+((kn_{c})^{2}+(k-1)^{2})(\frac{\sigma_{X}}{X})^{2}+(1+(kn_{c})^{2})(\frac{\sigma_{\rho}}{\rho})^{2}},\label{eq:9}
\end{equation}

where $\sigma_{X}$ and $\sigma_{\rho}$ are the uncertainties corresponding to the total length $X$ and mean density $\rho$, respectively. Note that the first term in Eq.(\ref{eq:9}) depends only on the statistics. In the next section we estimate the other terms in this equation. 

To apply these equations for $\Theta \ne 0$, the differential energy intensity for a given angle should also be parameterized in the form given by Eq. (\ref{eq:1}).  Here, to describe the angular variation we used $\cos(\Theta)^{m(E)}$, so the differential intensity for a given $\Theta$ becomes: 

\begin{equation}
F(E,\Theta)=GE^{-n(E)} \cos(\Theta)^{m(E)},  \label{eq:10}
\end{equation}

Because of the poor quality of the experimental data at large angles, $m(E)$ was obtained by fitting CORSIKA\cite{CORSIKA} simulated data. The procedure was to simulate the data over the entire angular range and normalize the results to the data in the small angular region $\Theta\leq5^{\circ}$. This provided an educated guess for the polynomial fit to the large angular part. The obtained results of $m(E)$ at sea level were fitted by a polynomial function, resulting in $m(E)=1.53-0.0484E+0.000545E^{2}-2.79 \times 10^{-6}E^{3}+5.11 \times 10^{-9}E^{4}$, where $E$ is expressed in $GeV$. This is represented in Fig 4 (solid line) together with the experimental data. As can be seen, the obtained results agree with the data within the experimental errors.

To present Eq (\ref{eq:10}) in the form of Eq (\ref{eq:1}) we introduced a function $g(\Theta) = G \cos(\Theta)^{m(E=10 GeV)}$ to describe the angular dependence of $G$, and $\eta(E,\Theta)$ for the spectral index, which becomes $n(E)$ at $\Theta = 0$.

\begin{equation}
F(E,\Theta)=g(\Theta)E^{-\eta(E,\Theta)},  \label{eq:11}
\end{equation}
For a given value of $\Theta$, approximate analytic calculations can also be carried out using a constant spectral index $\eta_c(\Theta)$, which likewise becomes $n_c$ at $\Theta = 0$. The value of $\eta_c(\Theta)$ would depend on the zenith angle, because of the additional $\cos(\Theta)^{m(E)}$ dependence. 

As shown in Fig 2, within the angular range $\Theta = 0^{o} - 60^{o}$, $\eta_c(\Theta)$ varies between $2.687$ and $2.53$, respectively. Also, within a $~10\%$ accuracy, at $\Theta = 60^{o}$ the approximation is appropriate for energies larger than $25$ $GeV$.
The decrease in $\eta_c(\Theta)$ relative to the vertical direction can reach $5-6\%$ and almost twice in $g(\Theta)$ parameter for  $\Theta = 60^{o}$. For $\eta_c(\Theta)$ a simple parameterization $\eta_c (\Theta)= n_c \cos(\theta)^{0.0907}$ can be used.
Another problem is that most of the available data compilations correspond to sea level measurements, while for other altitudes the experimental data is very scarce.
For small altitude changes ($h < 1000$ m, where $h$ in $m$ measured from the sea level), the variation in the shape of vertical differential energy spectrum for $E_{L} >10$ $GeV$ can be neglected. This is because the contribution of muon decay for those energies is negligible \cite{Nagamine}. Thus the altitude correction to the differential intensity can be applied for the normalization coefficient in Eq (\ref{eq:1}), using the exponential expression proposed in Ref. \cite{Hebecker,Belotti} as:

\begin{equation}
G_{h} = Gexp(h/h_{0}(E_{L}=10 GeV)),  \label{eq:12}
\end{equation}
Where $G_{h}$ is the normalization coefficient for the altitude $h$ and $h_{0}$  is a function of $E_{L}$ defined in \cite{Hebecker}. For large altitudes (for example $h = 2300$ m), the $m(E)$ obtained by fitting CORSIKA simulation data is shown in Fig 4 (dashed line). From the figure one can observe a difference between the $m(E)$ values obtained for the sea and Mexico City levels. So, for large altitudes, the value of $\eta_c(\Theta)$ and $g(\Theta)$ should be corrected.

\section{Uncertainties}

The second term in Eq.(\ref{eq:9}) is the uncertainty of the muon differential energy spectrum $F$. The uncertainty associated with the vertical differential energy spectrum has already been discussed in Section 2. Concerning $\cos(\Theta)^{m(E)}$, Eq.(\ref{eq:10}) provides good results at sea level, better than $5$\%, within $0^{\circ}\leq\Theta\leq40^{\circ}$. As illustrated in Fig. 5, beyond that angular region there is less data, and the corresponding error bars are bigger. For larger angles, the agreement with the experiment is $\approx 10\%-20\%$ for the energy interval $E < 100$ $GeV$ (Fig. 6).

Thus, for non-sea-level applications we also relied on $CORSIKA$ simulations to estimate the altitude behavior of the muon intensity. For altitudes $h\leq2300$ $m$ the accuracy of Eq. (\ref{eq:12}) has been checked using simulation data resulting in values better than $\leq5$ \%.

The third term in Eq.(\ref{eq:9}) represents the geometrical uncertainties, associated with the description of the external shape of the investigated volume, as well as those associated with the detector location. They define the muon path length $X$, which is known with some uncertainty $\sigma_{X}$. The $\Delta L$ determination is
affected by these geometrical uncertainties. However, the location of the detector only contributes to the systematic errors, introducing a fixed displacement but with no fluctuating character.

The last term in Eq.(\ref{eq:9}) represents the uncertainty in the internal density distribution. The simulation requires the best possible knowledge of the density and chemical composition of the materials contained in the volume of interest. This can be achieved by sampling the volume along some significant direction. When this procedure is carried on systematically,
one can define a sampling length $X^{samp}$, and estimate the density uncertainty $\sigma_{\rho}^{samp}$. Yet, we don't expect randomly distributed density variations to be important, because over a long $X$ the uncertainty in the mean value of the density depends on $X$, i.e., $\sigma_{\rho}(X)=\sigma_{\rho}^{samp}/\sqrt{X/X^{samp}}$.

Finally, there are a number of smaller factors known to affect the cosmic muon intensity ($E > 10 $ $GeV$), such as the Solar modulation, the Earth position and even the local atmospheric conditions, temperature and pressure. The importance of these effects has been reported by some authors  \cite{Hebecker,Grieder,Gawin} to be of the order $1 \%$.

\section{Limitations}

With the uncertainties described in the previous section, now we can estimate the minimum heterogeneity size as the uncertainty of $X$ (assuming negligible statistical error), using:

\begin{equation}
\Delta X_{min}  \approx  (X/k)\sqrt{(k\sigma_{X}/X)^{2}+
(\sigma_{\rho}(X)/\rho)^{2}+(\sigma_{F}/{F})^{2}},\label{eq:13}
\end{equation}

One factor not yet included in these estimations is the possibility that the heterogeneity of interest may not be a simple localized density excess, or defect, relative to the mean density calculated along a given muon trajectory, but a combination of them, reducing the sensitivity of the method. To quantify this, let us define the $\Delta X_{1}$ and $\Delta X_{2}$ as the total lengths corresponding to the densities $\rho_{1}$ and $\rho_{2}$, respectively, where $\rho_{1}<\rho<\rho_{2}$. Taking into account that $\xi  \propto \Delta L = \rho_{1}\Delta X_{1}+\rho_{2}\Delta X_{2}-\rho(\Delta X_{1}+\Delta X_{2})$ Eq. (\ref{eq:5}), one can conclude that $\xi$ is reduced when $\rho$ approaches $(\rho_{1}\Delta X_{1}+\rho_{2}\Delta X_{2})/(\Delta X_{1}+\Delta X_{2})$

\section{PSF and coordinate reconstruction uncertainty}

The effects of muon multiple scattering are best studied using Monte Carlo simulations. For this purpose we use the simulation package GEANT4 \cite{GEANT} which allows us to closely reproduce the relevant physical processes. An important parameter for heterogeneity localization is the coordinate reconstruction uncertainty ($\sigma_{r}$), associated with the reconstructed  muons location at a given density length distance from the detector ($L_{d}$). The knowledge of the coordinate transform (displacement) uncertainty ($\sigma_{p}$) is important for the optimization of experimental equipment. For example, if the detector size is less than $4\sigma_{p}$, then important information for image reconstruction will be lost. The relation between the object lateral size and $\sigma_{p}$ is also important. If the object's lateral size is comparable with $\sigma_{p}$, then one should expect a decrease in $\xi$.

The above-mentioned parameters depend not only on $L_{d}$ but also on the total thickness of the absorber for the given direction, because of the shift in the corresponding energy region, resulting a in a reduction of the fraction of low energy muons having large multiple scattering.
 
The $L_{d}$ dependence of the uncertainties $\sigma_{r}$ and $\sigma_{p}$ will be estimated within the range $2-16 kg/cm^{2}$ and for a total density length of $16 kg/cm^{2}$, which is expected to be the maximum thicknesses of our trial experiment \cite{Alfaro}(see next section). To estimate $\sigma_{r}$ and $\sigma_{p}$, a vertical beam of muons with the energy distribution given by Eq (\ref{eq:1}) was simulated, passing through different matter thicknesses, with the mean density $\rho\approx{2\times 10^{-3} kg/cm^{3}}$ \cite{Alfaro} and the standard soil material \cite{Chemistry}.  The penetrated muons were detected by the detector located below the matter volume having an ideal resolution.

The value of $\sigma_{p}$ is estimated as the standard deviation of the distribution of coordinate displacements on the detector surface. The back-projected coordinate distribution on the volume surface is obtained using the detected track information and $\sigma_{r}$ is estimated as the standard deviation of the corresponding distribution.

The results of $\sigma_{r}$ and $\sigma_{p}$ are presented in Fig. 7. Hence mentioned by Alvarez et al. \cite{Alvarez}, the use of low energy cuts on the detected muons does not significantly improve the resolution parameters. Both parameters have approximately linear dependence on the density length. In the figure the parameters for the case when $L_{d}=L$ are also presented, which clearly demonstrate the effect of total thickness.

In Ref. \cite{Grabski2} we showed that these simulations for multiple scattering angle with large energy cuts, and for different density lengths, are in agreement with the di-muon angular distribution in underground measurements by the L3C Collaboration \cite{Achard}.

\section{Heterogeneity simulation}

The purpose now is to validate the obtained approximate relations using realistic muon energy distributions based on Eq (\ref{eq:10}), and including effects of multiple scattering. To illustrate this we use GEANT4, taking the Pyramid of the Sun in Teotihuacan, Mexico, as a model of the matter volume. The complex geometrical shape of this monument is shown in Fig 8.

Two simulations using Eq. (\ref{eq:10}) have been performed with, and without, internal structures in the volume of the pyramid. For each simulation $5\times10^{7}$ events for zenith angle $0-50$ were generated.  As model heterogeneities we considered two empty volumes (see Fig 8). A chamber having $xyz$ dimensions ($16\times5\times4 m^{3}$), which are much larger than $\sigma_{p}$, so that multiple scattering will not strongly affect its detection. The other heterogeneity considered is a $60\times2\times1m^{3}$ tunnel, which is twice longer than an existing modern structure known as the Smith Tunnel. In this case the width of the tunnel is approximately the triple of $\sigma_{p}$ and one should expect a lower value for $\xi$.  The detector sensitive area of $1\times1m^{2}$ is also slightly smaller than $3 \sigma_{p}$, and will introduce some decrease of $\xi$.

The results of the Monte Carlo simulations for the detected muons, with hypothetical empty volumes are presented in Fig. 9 showing a plot of the projection angles $\Theta_{x}$ and $\Theta_{z}$ where $x$ and $z$ are the coordinates in the horizontal plane. The chamber is clearly seen, while a hint of the tunnel can also be observed. To make the latter more visible, the $\xi$ values obtained from a background subtraction are presented on Figs. 10-11 showing plots of projection angles $\Theta_{x}$ and $\Theta_{z}$. 

The value of the parameter $\xi$ determined in the center of the volume (see Fig. 10), is in good agreement with estimations using Eq (\ref{eq:5}) giving $\xi_{est} \approx 23$, $\xi_{sim} \approx 22$ (where $\xi_{est}$ and $\xi_{sim}$ are the estimated and the simulated values respectively). The estimated values of $\xi$ for the tunnel are expected to lie within $2-5$ because of the variation of parameter $L$  along the tunnel. From the one dimensional plot (Fig 11a) one can see that the obtained results lie in the interval of the expected values.

As one may expect, the summing of the signal bins along the tunnel will improve the signal to noise ratio (see Fig 11b). So for thin heterogeneities with large lateral sizes it is not necessary to collect large statistics to become detectable ($\xi\geq 5$ \cite{Bityukov} suggested by formula (\ref{eq:5})). So the approximate estimations for the statistics are good enough if the object's minimal lateral size and the detector size are larger than ($3-4$) $\sigma_{p}$.

\section{Conclusions}

The muon intensity attenuation method to detect heterogeneities in matter volumes for the density length region of $5-50 kg/cm^{2}$ has been analyzed. Approximate analytical expressions for the signal to noise ratio and for the statistics collection time to achieve a given signal to noise ratio have been proposed. The energy integral intensity measurement is shown to be preferable than the differential energy intensity for the density length larger $5 kg/cm^{2}$.  The coordinate reconstruction and transfer uncertainties have been estimated using Monte Carlo simulations for different thicknesses of the absorber. They show a linear dependence on the distance from the detector and a smaller dependence on the muon low energy cuts. The minimum detectable cavity size has been shown to depend on the thickness of the matter volume along the muon track.

\section{Acknowledgments}

We thank D. Heck, T. Hebbeker, C. Timerman and M. Zazian for discussions on the CORSIKA results.

\newpage 
%
\begin{figure}
\includegraphics[scale=1.3]{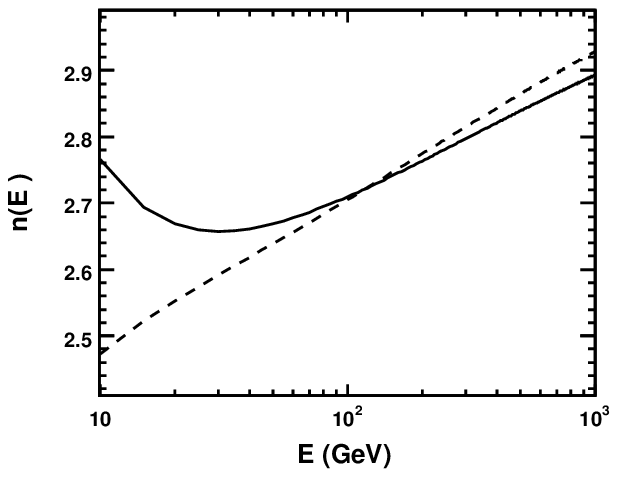}
\caption{Energy dependence of spectral index $n(E)$: The solid line using parameterization \cite{Hebecker} and the dashed line using parameterization \cite{CERN}.}
\end{figure}
\begin{figure}
\includegraphics[scale=1.3]{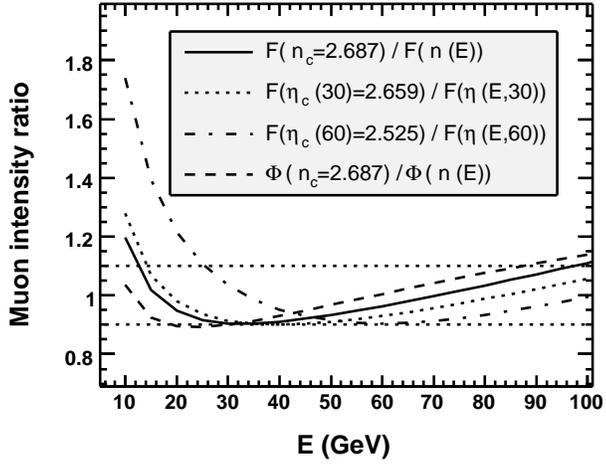}
\caption{Energy dependence of ratios: $F(n=constant)/F(n(E))$ and $\Phi(n=constant)/\Phi(n(E))$ for three different zenith angles. The horizontal lines demonstrate 10\% level of agreement.}
\end{figure}
\begin{figure}
\includegraphics[scale=1.2]{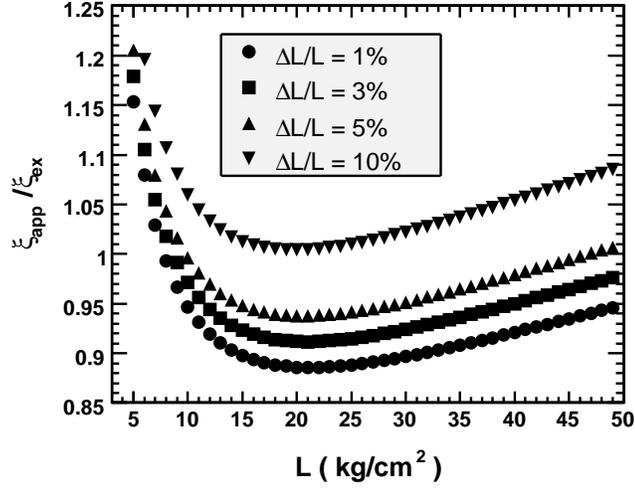}
\caption{$\xi$ ratios (approximate/exact) dependent on $L$ for different $\Delta L/L$ values.}
\end{figure}
\begin{figure}
\includegraphics[scale=1.2]{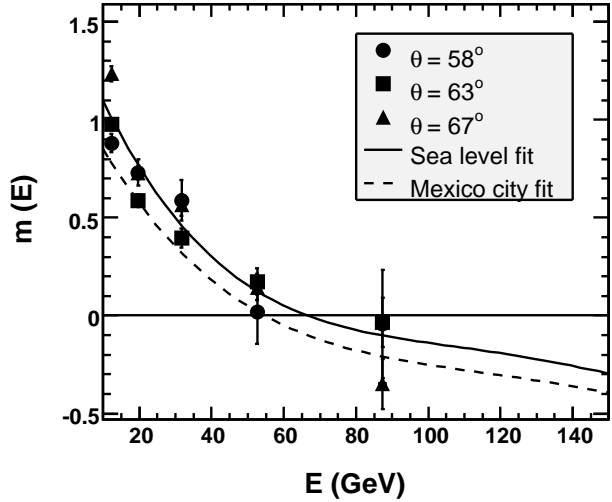}
\caption{The energy dependence of the power index $m$. The solid line is our fit based on CORSIKA simulated data at sea level, the dashed - the same at Mexico City altitude and points, the experimental data for three different angles at sea level \cite{Huston}.}
\end{figure}
\begin{figure}
\includegraphics[scale=1.2]{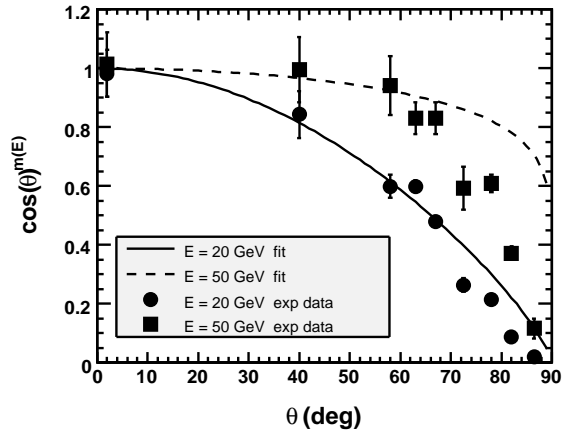}
\caption{Available\cite{Huston,Agrawal} large angle experimental zenith angular distributions 
for two representative energy values are shown as full circles and full squares, compared with the corresponding predictions of our fit (continuous and dashed, respectively)}.
\end{figure}
\begin{figure}
\includegraphics[scale=1.2]{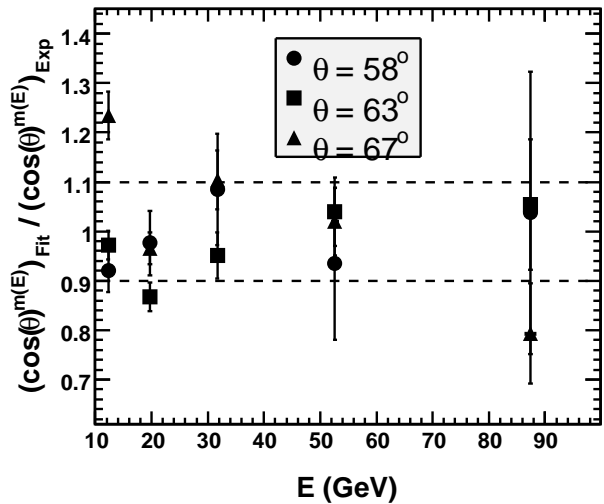}
\caption{Experimental data\cite{Huston} fit ratio for different zenith angles. The horizontal lines demonstrate 10\% level of agreement.}
\end{figure}
\begin{figure}
\includegraphics[scale=1.2]{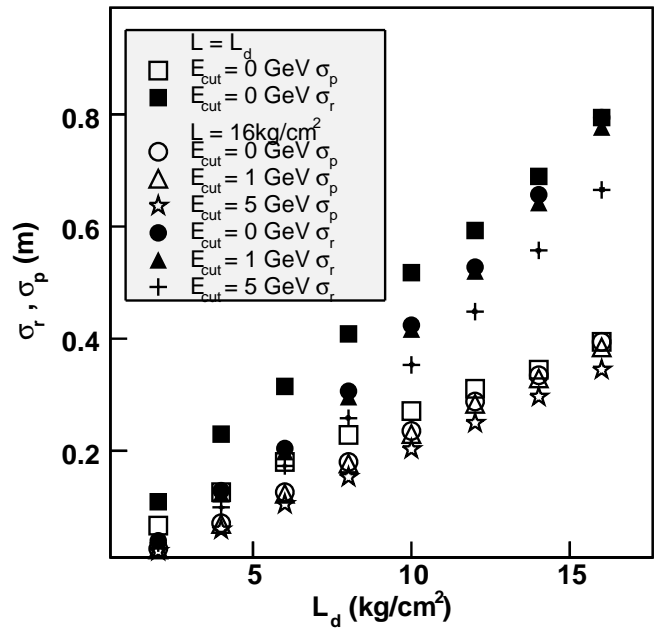}
\caption{Dependence of the $\sigma_{C}$(full symbols) and $\sigma_{P}$ (open symbols) on $L_{d}$ for different low energy cuts of the detected muons.}
\end{figure}
\begin{figure}
\includegraphics[scale=0.9]{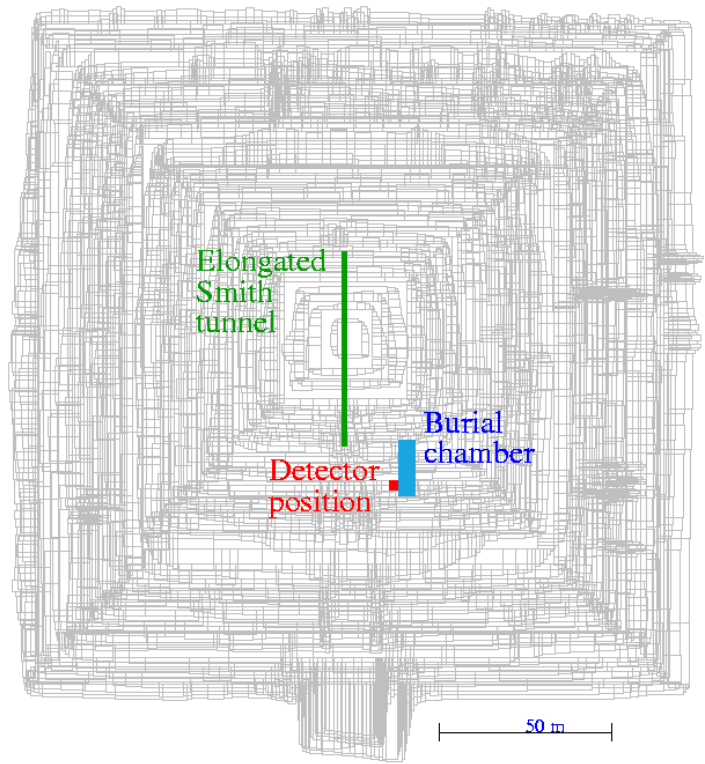}
\caption{Simulation setup}
\end{figure}
\begin{figure}
\includegraphics[scale=0.6]{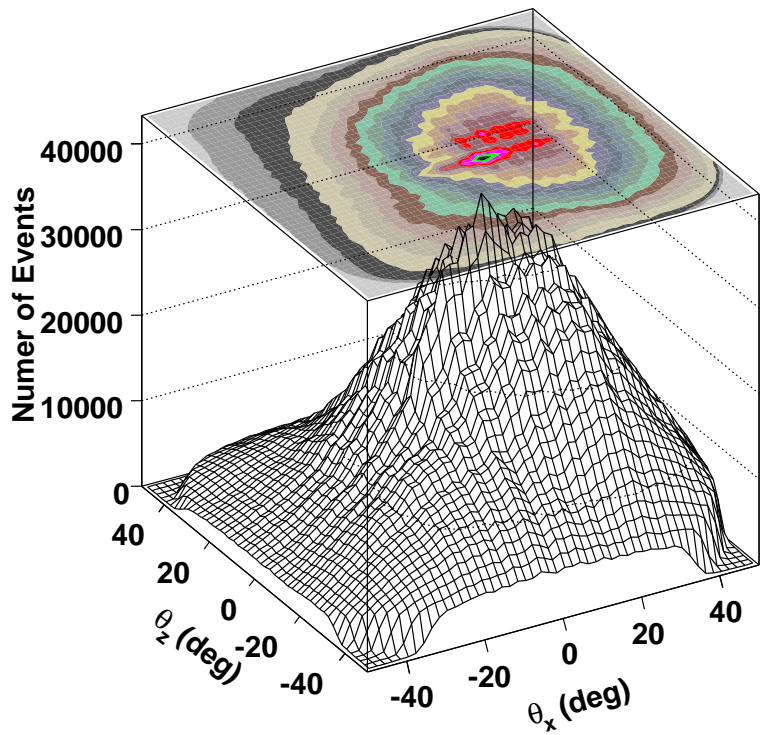}
\caption{Distribution of detected muons as a function of projection angles
considering that the elongated Smith tunnel and the hypothetical burial chamber are present. Layers of the pyramid and burial chamber (see text) are seen.}
\end{figure}
\begin{figure}
\includegraphics[scale=0.6]{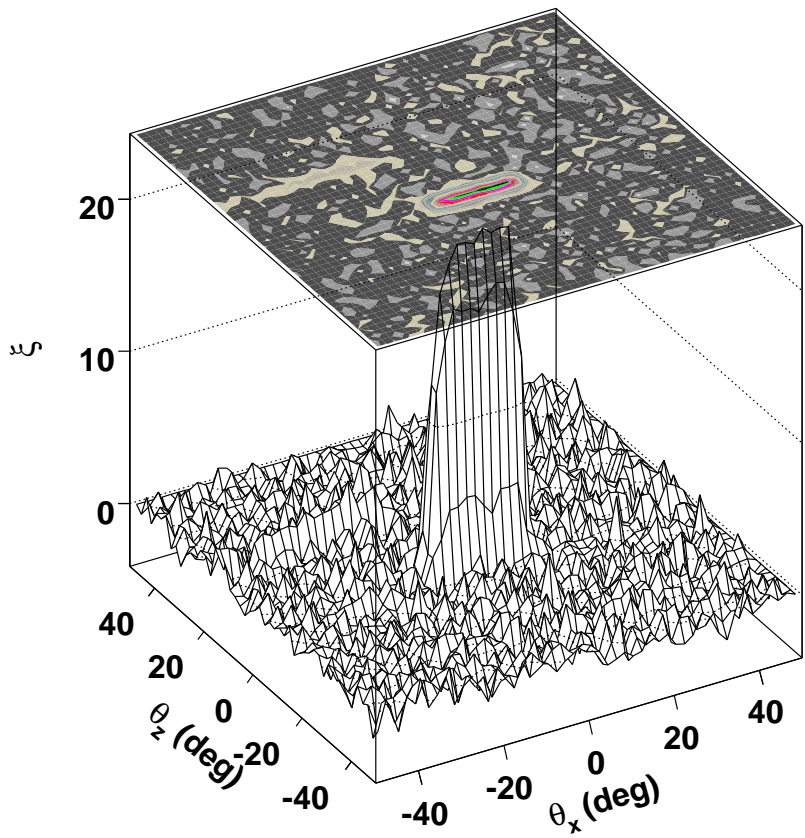}
\caption{Distribution of the sensitivity $\xi$ dependence on the projection
angles. The tunnel is seen as a bright short strip on the top. The big
chamber (see text) is seen very well.}
\end{figure}
\begin{figure}
\includegraphics[scale=1.2]{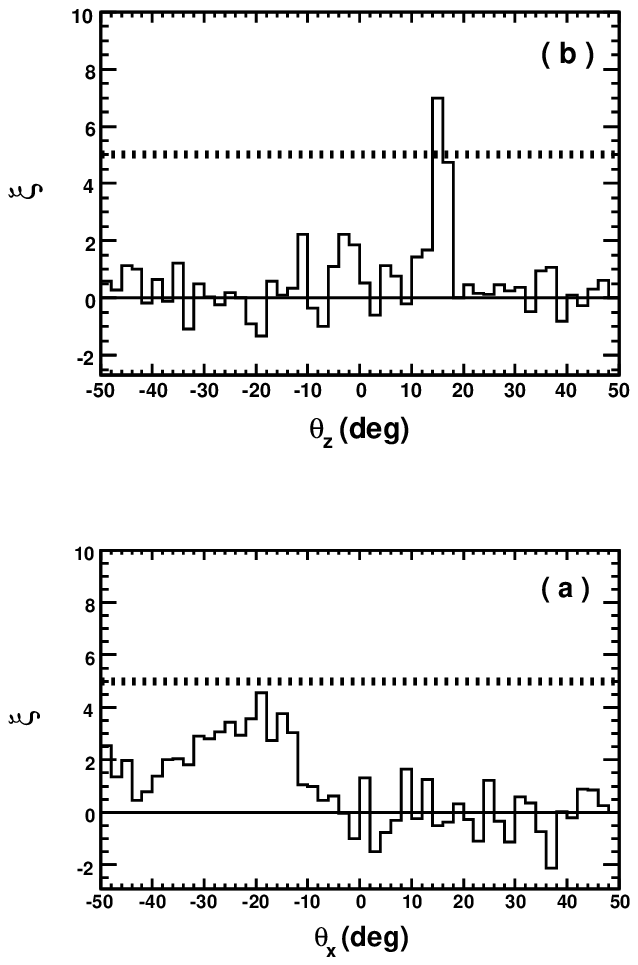}
\caption{Distribution of the parameter $\xi$ in dependence on the projection
angles $\Theta_{x}$ (a) and $\Theta_{z}$ (b).}
\end{figure}
\end{document}